# Electron Acoustic Solitary Structures in Nanoparticle Doped Semiconductor Quantum Plasma

**Abhishek Yadav and Punit Kumar***

*Department of Physics, University of Lucknow, Lucknow-226007, India*

*Email: kumar_punit@lkouniv.ac.in*

## ABSTRACT

The study of electron acoustic waves (EAWs) and their associated solitary structure in semiconductor quantum plasma doped with nanoparticle clusters have been carried out. The system consists of cold and hot electrons, holes, and stationary ions. The theory has been built using the quantum hydrodynamic (QHD) model. The dispersion relation for EAWs has been set up. To explore nonlinear behaviour, the perturbation technique has been applied, leading to the Korteweg de Vries (KdV) equation. The analysis demonstrates that quantum effects stabilize wave propagation at higher frequencies while presence of nanoparticles strongly influence wave dispersion at higher frequencies, resulting in greater dispersion. Nonlinear analysis shows that solitons in quantum plasma attain higher amplitudes and broader structures due to quantum effects and with the inclusion of nanoparticles.

1. **Introduction**

Quantum plasma physics [1-3] has a long and diverse tradition, with growing interest in its applications in modern technology and nanoscale electronic devices. Numerous theoretical and experimental studies on this subject exist in the literature, covering areas such as solid state physics, microelectronics [4], nanoparticles, metal nanostructures [5], quantum wells, quantum wires, quantum dots [6], ultracold plasmas [7], carbon nanotubes, and quantum diodes [8]. Quantum plasmas also play a key role in electron transport in metals and charge carrier dynamics in semiconductors. Research in semiconductor quantum plasma has steadily increased over the past decade, driven by unresolved challenges in charge transport due to the miniaturization of electronic components in micro and nanoelectronics. In semiconductor quantum plasmas, charge carriers obey the Fermi-Dirac distribution instead of the Maxwell-Boltzmann distribution due to the Pauli exclusion principle for fermions. In modern miniature semiconductor structures, the interparticle distances are comparable to the de Broglie thermal wavelengths of electrons and holes. As a result, quantum mechanical effects significantly influence the behaviour of future electronic components. For semiconductor quantum devices working with the electrons and holes in nanoscale sizes such as quantum wells and quantum dots, it is essential to understand and investigate thoroughly the quantum mechanical effects on the dynamics of the charged carriers.

The study of propagation of plasma waves in quantum semiconductor plasmas is of significant interest. Several studies have explored different aspects of this phenomenon. The propagation of surface electromagnetic waves in a magnetized quantum electron hole semiconductor (QEHS) plasma has been studied [9], along with the propagation of nonlinear acoustic waves [10], and the electron-hole two stream instability [11] in such plasmas. Additionally, the behaviour and instability of electrostatic perturbations in a QEHS plasma driven by an energetic electron beam have been examined [12]. Research has also focused on the quantum effects arising in linear and nonlinear quantum electrostatic acoustic waves [13] and the modulational instability of quantum acoustic waves [14]. Moreover, the nonlinear dispersion of waves in QEHS plasmas has been reported [15], and the impact of the tunneling of degenerate plasma species through the Bohm potential barrier on nonlinear acoustic wave propagation has been investigated [16].

The flow dynamics of charge carriers in semiconductor plasma present many unresolved challenges. One of the thrust areas in the perspective of flow dynamics relates to the study of Electron Acoustic Waves (EAWs). EAWs are high frequency dispersive plasma waves with frequencies much higher than the ion frequency. Since the EAW frequency exceeds the ion plasma frequency, ion dynamics does not influence the EAWs, and thus, ions can be considered stationary, with their charge uniformly distributed throughout the plasma. The propagation of EAWs has been studied in both magnetized [17, 18] and unmagnetized plasmas [19, 20]. Experimental observations of EAWs have been reported in magnetized and unmagnetized one dimensional, collisionless plasmas consisting of three components [21, 22].

EAWs do exhibit soliton formation and have been extensively studied both theoretically and experimentally [23-25]. The evolution of small amplitude EAWs is typically described by nonlinear equations such as the KdV, the Zakharov Kuznetsov (ZK), and the nonlinear Schrödinger equation [26-30]. Studies on the nonlinear propagation of EAWs have gained importance in explaining observations from both laboratory and space plasmas [31-33]. Various theoretical investigations have focused on electron acoustic (EA) solitons [34-36]. In recent years, the nonlinear propagation of EAWs in quantum plasmas with an unbounded planar geometry has been explored [37, 38]. Dispersion properties of EAWs and the formation of bright and dark EA solitons in unmagnetized quantum plasmas have been analyzed [39]. Oblique modulational instability of EAWs in quantum plasmas has been investigated [40]. Recent studies have also examined the linear and nonlinear properties of obliquely propagating EA solitary waves in a two electron temperature quantum magnetoplasma [41].

In the context of semiconductors, doping is an important phenomenon responsible for the performance of solid state devices. The development of high frequency solid state devices has gained great attention due to their potential applications in radio astronomy, industry, and defence [42]. Here extremely small feature sizes are required, but to achieve such small sizes, carrier transit time always imposes limitations and these limitations could not be overcome by using conventional approaches. One possible solution to overcome this limitation is to incorporate the properties of nanoparticles (NPs) into existing plasma media. Nanoparticle (NP) clusters exhibit properties that lie between those of macroscopic solids and atomic or molecular systems due to their size dependent characteristics. Their presence in a medium can significantly alter the bulk material's properties. NP clusters also possess unique optical properties, making them valuable for next generation optical sensors and nanodevices [43]. Modifications in the thermal properties of bulk semiconductors when embedded with metallic nanoparticles have been reported [44-47]. Numerous studies have explored the growth mechanisms and significant modifications in the physical and chemical properties of semiconductors [48-51]. NPs play a crucial role in crystal growth in organic semiconductors by altering charge carrier mobility [52, 53]. Extensive research has been conducted on the fabrication and characterization of both inorganic and organic semiconductors. However, the theoretical understanding of EAWs and their modeling by incorporating quantum effects in a semiconducting plasma system embedded with nanoparticles, remains unexplored.

The present work focuses on examining the electron-acoustic waves in a semiconductor quantum plasma doped with NP clusters. The system consists of low temperature, inertial cold electrons and high temperature, inertialess hot electrons along with holes and stationary ions. The primary objective is to investigate the influence of quantum mechanical effects and NPs on the structure of EA solitary waves, which have remained relatively unexplored in previous studies. To achieve this, QHD model and the KdV approach have been employed to establish the dispersion relation and soliton solutions, highlighting the impact of nanoparticle doping on the propagation characteristics of EAWs. The QHD model provides a self consistent approach to studying quantum plasma systems, incorporating quantum corrections in the form of the Bohm potential and Fermi pressure. Unlike kinetic models such as the Wigner Poisson system, QHD model offers numerical efficiency, simpler boundary condition implementation, and direct access to macroscopic variables like momentum and energy. QHD model is very useful to study the short scale collective phenomena, including waves, instabilities, and nonlinear interactions in dense plasmas. The KdV equation has been established to describe the Solitary wave propagation in unmagnetized plasmas without the dissipation and geometry distortion. This study concentrates on a heavily doped extrinsic n-InSb semiconductor as the medium. Such a study has not been reported in literature so far, and the theory is equally applicable to other similar semiconductors as well.

In Section 2, the fundamental theoretical formulation required for the study has been built up. In Section 3, the dispersion relation has been derived. In section 4, the KdV equation has been established and finally in section 5, the summary and a discussion of the results obtained has been presented.

## 2. Formalism

We consider a semiconductor quantum plasma composed of two distinct populations of electrons, low-temperature inertial cold electrons, and high-temperature inertialess hot electrons with holes and stationary ions forming a neutralizing background. We consider, the plasma to be doped with nanoparticle clusters, characterized by a number density $(N)$, electron density $(n_{0n})$, and radius $(r)$. The cold electrons, due to their low temperature and reduced mobility, provide the inertial contribution to the dynamics, while the hot electrons, with their higher temperature and greater mobility, offer the restoring force. So, in the momentum equation for hot electrons, we consider inertia term to be zero. The phase speed of the EAW lies in the range $v_{Fce} \ll \omega/k \ll v_{Fhe}$, where $v_{Fce}$ and $v_{Fhe}$ are the Fermi velocities of cold and hot electrons, respectively. Since $n_{0ce} \ll n_{0he}$ holds for EAW, which implies that $T_{Fce} \ll T_{Fhe}$ in quantum plasmas, therefore the Fermi pressure due to cold electrons can be ignored in comparison to the hot electrons in the model.

The basic set of governing QHD fluid eqs. comprise of continuity, momentum and Poisson eqs. as given below,

$$\frac{\partial n_\alpha}{\partial t} + \frac{\partial}{\partial x}(n_\alpha \vec{v}_\alpha) = 0, \tag{1a}$$

$$\frac{\partial \vec{v}_\alpha}{\partial t} + \vec{v}_\alpha \frac{\partial \vec{v}_\alpha}{\partial x} = \frac{q_\alpha}{m_\alpha}\frac{\partial \phi}{\partial x} - \frac{1}{m_\alpha n_\alpha}\frac{\partial P_{F\alpha}}{\partial x} + \frac{\hbar^2}{2m_\alpha^2}\frac{\partial}{\partial x}\left(\frac{1}{\sqrt{n_\alpha}}\frac{\partial^2}{\partial x^2}\sqrt{n_\alpha}\right), \tag{1b}$$

$$\varepsilon_0 \frac{\partial^2 \phi}{\partial x^2} = \sum_{\alpha=ce,he,h,n} q_\alpha n_\alpha. \tag{1c}$$

Eq. (1a) is the continuity equation for plasma species $\alpha$, where $\alpha$ represents cold electrons $(ce)$, hot electrons $(he)$, holes $(h)$ and electrons $(n)$ inside the nanoparticles, $n_\alpha$ is the particle density and, $v_\alpha$ is the velocity of the respective particle. Eqn. (1b) is the to momentum eqn. for plasma species $\alpha$. The second term, on the left hand side of above eq. (1b) is the convective time derivative of the velocity. The first term on RHS is the force due to the electrostatic potential $(\phi)$, $q_\alpha$ is the charge on the particle and $m_\alpha$ is the particle's effective mass. The second term is the force due to Fermi pressure $P_{F\alpha}\left(=m_\alpha v_{F\alpha}^2 n_\alpha^3/3n_{0\alpha}^2\right)$, where $v_F\left(=\sqrt{2k_B T_F/m_\alpha}\right)$, is the Fermi velocity. The third term is the force of the quantum Bohm potential arising from quantum corrections in density fluctuations and effects the phase and group velocities in semiconductor plasma, where h is the reduced Planck's constant. Eq. (1c) is the Poisson's equation for plasma species $\alpha$, where $\varepsilon_0$ is the electrical permittivity of free space.

The equations describing the motion of respective plasma species are,

$$\frac{\partial \vec{v}_{ce}}{\partial t} + \vec{v}_{ce} \frac{\partial \vec{v}_{ce}}{\partial x} = \frac{e}{m_{ce}} \frac{\partial \phi}{\partial x} + \frac{h^2}{2m_{ce}^2} \frac{\partial}{\partial x} \left( \frac{1}{\sqrt{n_{ce}}} \frac{\partial^2}{\partial x^2} \sqrt{n_{ce}} \right), \tag{2}$$

$$0 = \frac{e}{m_{he}} \frac{\partial \phi}{\partial x} - \frac{1}{m_{he} n_{he}} \frac{\partial P_{Fhe}}{\partial x} + \frac{h^2}{2m_{he}^2} \frac{\partial}{\partial x} \left( \frac{1}{\sqrt{n_{he}}} \frac{\partial^2}{\partial x^2} \sqrt{n_{he}} \right), \tag{3}$$

$$0 = -\frac{e}{m_h} \frac{\partial \phi}{\partial x} - \frac{1}{m_h n_h} \frac{\partial P_{Fh}}{\partial x} + \frac{h^2}{2m_h^2} \frac{\partial}{\partial x} \left( \frac{1}{\sqrt{n_h}} \frac{\partial^2}{\partial x^2} \sqrt{n_h} \right), \tag{4}$$

$$\frac{\partial^2 \Delta}{\partial t^2} + \frac{\omega_{pn}^2}{3} \Delta = \frac{e}{m_e} \frac{\partial \phi}{\partial x} - \frac{1}{m_e n_n} \frac{\partial P_{Fn}}{\partial x} + \frac{h^2}{2m_e^2} \frac{\partial}{\partial x} \left( \frac{1}{\sqrt{n_n}} \frac{\partial^2}{\partial x^2} \sqrt{n_n} \right), \tag{5}$$

and

$$\frac{\partial^2 \phi}{\partial x^2} = \frac{e}{\varepsilon_0} \left( n_{ce} + n_{he} + Z_n n_n - n_h - Z_i n_{0i} \right). \tag{6}$$

Eqs. (2), (3), (4) and (5) correspond to momentum equation for cold electrons $(ce)$, hot electrons $(he)$, holes $(h)$ and electrons inside the nanoparticles $(n)$ respectively. Eqn. (6) is the Poisson's equation where, $Z_n$ is the charge state of the nanoparticles in the system denoting the effective number of elementary charges carried by each nanoparticle, $Z_i$ is the charge state of ions in the system and, the subscript $i$ in the Poisson's equation is for ions.

### 3. Dispersion relation

In order to establish the linear dispersion relation for the quantum electron acoustic mode, we employ the perturbation technique and we make the following perturbation expansion for the field quantities about their equilibrium values,

$$\begin{pmatrix} n_{he} \\ n_{ce} \\ n_h \\ n_n \\ v_c \\ v_n \\ \phi \end{pmatrix} = \begin{pmatrix} n_{0he} \\ n_{0ce} \\ n_{0h} \\ n_{0n} \\ v_{0ce} \\ v_{0n} \\ \phi_0 \end{pmatrix} + \chi \begin{pmatrix} n_{he}^{(1)} \\ n_{ce}^{(1)} \\ n_h^{(1)} \\ n_n^{(1)} \\ v_{ce}^{(1)} \\ v_n^{(1)} \\ \phi^{(1)} \end{pmatrix} + \chi^2 \begin{pmatrix} n_{he}^{(2)} \\ n_{ce}^{(2)} \\ n_h^{(2)} \\ n_n^{(2)} \\ v_{ce}^{(2)} \\ v_n^{(2)} \\ \phi^{(2)} \end{pmatrix}. \tag{7}$$

where, $\chi$ is the strength of nonlinearity, superscripts represent the order of perturbation.

Substituting the perturbed quantities in eqs. (1(a)-6) and linearizing, assuming all the perturbed quantities to vary as $e^{i(kz-\omega t)}$, we get,

$$v_\alpha^{(1)} = \frac{\omega n_\alpha^{(1)}}{k n_{0\alpha}}, \tag{8}$$

$$v_{ce}^{(1)} = Q_C \frac{k^3}{\omega} n_{ce}^{(1)} - \frac{ke}{\omega m_e} \phi^{(1)}, \tag{9}$$

$$0 = \frac{e}{m_e} \phi^{(1)} - Q_{he} k^2 n_{he}^{(1)} - \frac{v_{Fhe}^2}{n_{0he}} n_{he}^{(1)}, \tag{10}$$

$$0 = -\frac{e}{m_h} \phi^{(1)} - Q_h k^2 n_h^{(1)} - \frac{v_{Fh}^2}{n_{0h}} n_h^{(1)}, \tag{11}$$

$$v_n^{(1)} = \frac{\left( \frac{e\omega k}{m_e} \phi^{(1)} - Q_n \omega k^3 n_n^{(1)} - \frac{v_{Fn}^2 \omega k}{n_{0n}} n_n^{(1)} \right)}{\left( \frac{\omega_{pn}^2}{3} - \omega^2 \right)}, \tag{12}$$

and

$$k^2\phi^{(1)} = -\frac{e}{\varepsilon_0}\left(n_c^{(1)} + n_{he}^{(1)} + Z_n n_n^{(1)} - n_h^{(1)}\right). \tag{13}$$

where, $Q_c\left(=\frac{\hbar^2}{4m_c^2 n_{0ce}}\right)$, $Q_{he}\left(=\frac{\hbar^2}{4m_{he}^2 n_{0he}}\right)$, $Q_h\left(=\frac{\hbar^2}{4m_h^2 n_{0h}}\right)$, and $Q_n\left(=\frac{\hbar^2}{4m_n^2 n_{0n}}\right)$ are the terms corresponding to Bohm potential of cold electrons, hot electrons, holes and electrons in nanoparticles respectively. $n_{0ce}$, $n_{0he}$, $n_{0h}$, and $n_{0n}$ are the equilibrium number density of cold electrons, hot electrons, holes and electrons in nanoparticles respectively.

Simultaneously solving eqs. (8-13), we arrive at the following dispersion relation of the EAW in the multispecies unmagnetized semiconductor quantum plasma,

$$k^2 = \left(\frac{\omega_{pce}^2}{\left(\frac{\omega^2}{k^2} - Q_c k^2 n_{0c}\right)} + \frac{Z_n \omega_{pn}^2}{\left(\frac{\omega^2}{k^2} - v_{Fn}^2 - Q_n k^2 n_{0n} - \frac{\omega_{pn}^2}{3k^2}\right)} - \frac{\omega_{phe}^2}{\left(Q_{he}k^2 n_{0he} + v_{Fhe}^2\right)} - \frac{\omega_{ph}^2}{\left(Q_h k^2 n_{0h} + v_{Fh}^2\right)}\right). \tag{14}$$

Above equation represents the dispersion relation for electron acoustic waves (EAWs) in a multispecies unmagnetized semiconductor quantum plasma. This shows, how the different species (cold electrons, hot electrons, holes and nanoparticles.) contribute to the propagation of electron acoustic waves in plasma. The term on left-hand side, is the square of the wavenumber, encapsulating the spatial properties of the wave. On the right-hand side, the first term, gives the contribution of cold electrons. The second term, accounts for the influence of nanoparticles where, the factor $Z_n$ is the the effective charge contribution of nanoparticles. The third term, describes the contribution of hot electrons. The fourth term, signifies the role of holes in the plasma. Here, $\omega_{p\alpha}\left(=\sqrt{n_{0\alpha}e^2/m_\alpha\varepsilon_0}\right)$ represents the plasma frequency of plasma species $\alpha$. Each term contributes uniquely to the overall dispersion relation, reflecting the interplay between different plasma species and the quantum effects that govern their dynamics.

Fig. 1 shows the variation of $\omega/\omega_p$ with $kc/\omega_p$ in semiconductor quantum plasma. The solid line shows the variation in quantum plasma, while the dashed line shows the trend in absence of quantum effects. In the low frequency region, both curves nearly coincide. As the frequency increases, the curve in absence of quantum effects begins to rise more steeply. A significant separation between the two curves is observed in the high-frequency region. The quantum curve shows a slower increase in wavenumber, indicating that quantum effects stabilizes wave propagation and prevent excessive growth in wave number at higher frequencies. This is due to the quantum correction terms (Bohm potential and Fermi pressure) which act as a stabilizing factor by introducing additional restoring forces in the plasma system leading to controlled increment in wavenumber.

Fig. 2 shows the variation of $\omega/\omega_p$ with $kc/\omega_p$ for different values of cold to hot electron density ratio. In the low frequency region, all three curves nearly overlap, showing minimal variation in wavenumber. In the high frequency region, the separation between the three curves increases, and the wavenumber rises more rapidly for lower cold to hot electron density ratios. This indicates that an increase in cold electron density leads to lower wavenumber growth, resulting in reduced wave dispersion. As cold electron density increases, collective motion cold electrons dominates. Even though cold electrons have lower individual velocities, their collective effect contributes to an increase in phase velocity and this leads to lower wavenumber growth.

Fig. 3 shows the variation of $\omega/\omega_p$ with $kc/\omega_p$ in semiconductor quantum plasma in presence and absence of nanoparticle cluster. The separation between the two curves is significant at high frequencies, and the wavenumber increases much more rapidly in the presence of nanoparticle clusters. This indicates that nanoparticles have a strong influence on wave dispersion at higher frequencies, resulting in greater dispersion because each nanoparticle introduce additional electrons increasing the overall plasma electron density that participate in plasma oscillations with their quantum pressure and quantum Bohm potential. At higher frequencies, quantum effects become more significant and alters the wave behavior leading to greater separation between the two curves which we have already seen in discussion of figure 1.

**4. Soliton Solution**

In order to study the nonlinear behaviour of electron acoustic wave (EAW), we introduce the stretched coordinates,

$$\zeta = \chi^{1/2}(x - \lambda t), \tag{15}$$

and

$$\tau = \chi^{3/2} t, \tag{16}$$

where, $\lambda$ is the phase velocity of wave. In this transformation, $x$ and $t$ are function of $\zeta$ and $\tau$ respectively, so partial derivatives with respect to $x$ and $t$ can be transformed into partial derivative in terms of $\zeta$ and $\tau$ as,

$$\frac{\partial}{\partial x} = \chi^{1/2} \frac{\partial}{\partial \zeta}, \tag{17}$$

$$\frac{\partial}{\partial t} = -\chi^{1/2} \lambda \frac{\partial}{\partial \zeta} + \chi^{3/2} \frac{\partial}{\partial \tau}, \tag{18}$$

$$\frac{\partial^2}{\partial x^2} = \chi \frac{\partial^2}{\partial \zeta^2}, \tag{19}$$

and

$$\frac{\partial^3}{\partial x^3} = \chi^{3/2} \frac{\partial^3}{\partial \zeta^3}. \tag{20}$$

Eqs. (1-6) in terms of $\zeta$ and $\tau$ can be written as,

$$0 = -\chi^{1/2} \lambda \frac{\partial n_\alpha}{\partial \zeta} + \chi^{3/2} \frac{\partial n_\alpha}{\partial \tau} + \chi^{1/2} \frac{\partial}{\partial \zeta}(n_\alpha v_\alpha), \tag{21}$$

$$-\chi^{1/2} \lambda \frac{\partial v_c}{\partial \zeta} + \chi^{3/2} \frac{\partial v_c}{\partial \tau} + \chi^{1/2} v_c \frac{\partial}{\partial \zeta} v_c = \frac{e}{m_e} \chi^{1/2} \frac{\partial \phi}{\partial \zeta} + Q_c \chi^{3/2} \frac{\partial^3 n_c}{\partial \zeta^3}, \tag{22}$$

$$0 = \frac{e}{m_e} \chi^{1/2} \frac{\partial \phi}{\partial \zeta} + Q_{he} \chi^{3/2} \frac{\partial^3 n_{he}}{\partial \zeta^3} - \frac{v_{Fhe}^2}{n_{0he}^2} n_{he} \chi^{1/2} \frac{\partial n_{he}}{\partial \zeta}, \tag{23}$$

$$0 = -\frac{e}{m_h} \chi^{1/2} \frac{\partial \phi}{\partial \zeta} + Q_h \chi^{3/2} \frac{\partial^3 n_h}{\partial \zeta^3} - \frac{v_{Fh}^2}{n_{0h}^2} n_h \chi^{1/2} \frac{\partial n_h}{\partial \zeta}, \tag{24}$$

$$-\chi^{1/2} \lambda \frac{\partial v_n}{\partial \zeta} + \chi^{3/2} \frac{\partial v_n}{\partial \tau} + \frac{\omega_{pn}^2}{3}V + \chi^{1/2} v_n \frac{\partial v_n}{\partial \zeta} = \frac{e}{m_e} \chi^{1/2} \frac{\partial \phi}{\partial \zeta} + Q_n \chi^{3/2} \frac{\partial^3 n_n}{\partial \zeta^3} - \frac{v_{Fn}^2}{n_{0n}^2} n_n \chi^{1/2} \frac{\partial n_n}{\partial \zeta}, \tag{25}$$

and

$$\chi \frac{\partial^2 \phi}{\partial \zeta^2} = \frac{e}{\varepsilon_0}(n_c + n_{he} + Z_n n_n - n_h - Z_i n_{0i}). \tag{26}$$

Substituting the perturbations defined by eq. (7) in eqs. (21) – (26), and equating the lowest order terms, we get

$$n_{0\alpha} \frac{\partial v_\alpha^{(1)}}{\partial \zeta} - \lambda \frac{\partial n_\alpha^{(1)}}{\partial \zeta} = 0, \tag{27}$$

$$\frac{e}{m_e}\frac{\partial \phi^{(1)}}{\partial \zeta} + \lambda \frac{\partial v_c^{(1)}}{\partial \zeta} = 0, \tag{28}$$

$$\frac{e}{m_e}\frac{\partial \phi^{(1)}}{\partial \zeta} - \frac{v_{Fhe}^2}{n_{0he}}\frac{\partial n_{he}^{(1)}}{\partial \zeta} = 0, \tag{29}$$

$$\frac{e}{m_h}\frac{\partial \phi^{(1)}}{\partial \zeta} - \frac{v_{Fh}^2}{n_{0h}}\frac{\partial n_h^{(1)}}{\partial \zeta} = 0, \tag{30}$$

$$\frac{e}{m_e}\frac{\partial \phi^{(1)}}{\partial \zeta} - \frac{v_{Fn}^2}{n_{0n}}\frac{\partial n_n^{(1)}}{\partial \zeta} + \lambda \frac{\partial v_n^{(1)}}{\partial \zeta} = 0, \tag{31}$$

and

$$\frac{e}{\varepsilon_0}\left(n_c^{(1)} + n_{he}^{(1)} + Z_n n_n^{(1)} - n_h^{(1)}\right) = 0. \tag{32}$$

Integrating and simplifying the above equations, yield

$$n_c^{(1)} = -\frac{en_{0c}}{m_e \lambda^2}\phi^{(1)}, \tag{33}$$

$$n_{he}^{(1)} = \frac{en_{0he}}{m_e v_{Fhe}^2}\phi^{(1)}, \tag{34}$$

$$n_h^{(1)} = -\frac{en_{0h}}{m_h v_{Fh}^2}\phi^{(1)}, \tag{35}$$

$$n_n^{(1)} = -\frac{en_{0n}}{m_e\left(\lambda^2 - v_{Fn}^2\right)}\phi^{(1)}, \tag{36}$$

and

$$\frac{e}{\varepsilon_0}\left(n_c^{(1)} + n_{he}^{(1)} + Z_n n_n^{(1)} - n_h^{(1)}\right) = 0. \tag{37}$$

The phase velocity is obtained by simultaneously solving the above eqs.,

$$\lambda = \left(\frac{n_{0c} + \frac{Z_n n_{0n}}{\left(1 - v_{Fn}^2\right)}}{\frac{n_{0he}}{v_{Fhe}^2} + \frac{n_{0h}}{v_{Fh}^2}}\right)^{1/2}. \tag{38}$$

Substituting the perturbations defined by eqs. (7) in eqs. (21) – (26), and collecting the higher order terms, we get

$$-\lambda \frac{\partial n_\alpha^{(2)}}{\partial \zeta} + \frac{\partial n_\alpha^{(1)}}{\partial \tau} + n_{o\alpha} \frac{\partial v_\alpha^{(2)}}{\partial \zeta} + \frac{\partial}{\partial \zeta}\left(n_\alpha^{(1)} v_\alpha^{(1)}\right) = 0, \tag{39}$$

$$-\lambda \frac{\partial v_c^{(2)}}{\partial \zeta} + \frac{\partial v_c^{(1)}}{\partial \tau} + v_c^{(1)} \frac{\partial v_c^{(1)}}{\partial \zeta} = \frac{e}{m_e} \frac{\partial \phi^{(2)}}{\partial \zeta} + Q_c \frac{\partial^3 n_c^{(1)}}{\partial \zeta^3}, \tag{40}$$

$$\frac{e}{m_e} \frac{\partial \phi^{(2)}}{\partial \zeta} - \frac{v_{Fhe}^2}{n_{0he}^2} \frac{\partial n_{he}^{(2)}}{\partial \zeta} - \frac{v_{Fhe}^2}{n_{0he}^2}\left(n_{he}^{(1)} \frac{\partial n_{he}^{(1)}}{\partial \zeta}\right) + Q_{he} \frac{\partial^3 n_{he}^{(1)}}{\partial \zeta^3} = 0, \tag{41}$$

$$Q_h \frac{\partial^3 n_h^{(1)}}{\partial \zeta^3} - \frac{e}{m_h} \frac{\partial \phi^{(2)}}{\partial \zeta} - \frac{v_{Fh}^2}{n_{0h}^2} \frac{\partial n_h^{(2)}}{\partial \zeta} - \frac{v_{Fh}^2}{n_{0h}^2}\left(n_h^{(1)} \frac{\partial n_h^{(1)}}{\partial \zeta}\right) = 0,$$

(42)

$$-\lambda \frac{\partial v_n^{(2)}}{\partial \zeta} + \frac{\partial v_n^{(1)}}{\partial \tau} + v_n^{(1)} \frac{\partial v_n^{(1)}}{\partial \zeta} = \frac{e}{m_e} \frac{\partial \phi^{(2)}}{\partial \zeta} + Q_n \frac{\partial^3 n_n^{(1)}}{\partial \zeta^3} - \frac{v_{Fn}^2}{n_{0n}^2} \frac{\partial n_n^{(2)}}{\partial \zeta} - \frac{v_{Fn}^2}{n_{0n}^2}\left(n_n^{(1)} \frac{\partial n_n^{(1)}}{\partial \zeta}\right), \tag{43}$$

and

$$\frac{\partial^2 \phi^{(1)}}{\partial \zeta^2} = \frac{e}{\varepsilon_0}\left(n_c^{(2)} + n_{he}^{(2)} + Z_n n_n^{(2)} - n_h^{(2)}\right). \tag{44}$$

Solving eqs. (39) – (44), we finally obtain

$$\frac{\partial \phi^{(1)}}{\partial \tau} + A_1 \phi^{(1)} \frac{\partial \phi^{(1)}}{\partial \zeta} + A_2 \frac{\partial^3 \phi^{(1)}}{\partial \zeta^3} = 0. \tag{45}$$

Eq. (45) is the required KdV equation, where the nonlinearity coefficient $A_1$ and the dispersive coefficient $A_2$ are,

$$A_1 = \frac{\left(\dfrac{e\omega_{phe}^2}{m_e v_{Fhe}^4} - \dfrac{3e\omega_{pc}^2}{m_e \lambda^4} - \dfrac{e\omega_{ph}^2}{m_h v_{Fh}^4} - \dfrac{4\pi r^3 N e \omega_{pn}^2 \left(3\lambda^2 + v_{Fn}^2\right)}{3 m_e \left(\lambda^2 - v_{Fn}^2\right)^3}\right)}{\dfrac{2}{\lambda}\left(\dfrac{\omega_{pc}}{\lambda}\right)^2 + \dfrac{8\pi r^3 N \lambda}{3}\left(\dfrac{\omega_{pn}}{\left(\lambda^2 - v_{Fn}^2\right)}\right)^2},$$

and, $\quad A_2 = \dfrac{\left(1 - \dfrac{n_{0c}\omega_{pc}^2 Q_c}{\lambda^4} - \dfrac{n_{0he}\omega_{phe}^2 Q_{he}}{v_{Fhe}^4} - \dfrac{n_{0h}\omega_{ph}^2 Q_h}{v_{Fh}^4} - \dfrac{4\pi r^3 N n_{0n} \omega_{pn}^2 Q_n}{3\left(\lambda^2 - v_{Fn}^2\right)^2}\right)}{\dfrac{2}{\lambda}\left(\dfrac{\omega_{pc}}{\lambda}\right)^2 + \dfrac{8\pi r^3 N \lambda}{3}\left(\dfrac{\omega_{pn}}{\left(\lambda^2 - v_{Fn}^2\right)}\right)^2}.$

To obtain a traveling wave solution to the KdV equation (45), we transform the stretched coordinates $\zeta$ and $\tau$ into one coordinate $\eta = \zeta - U\tau$, where $U$ is the constant speed of the solitary wave. Boundary conditions are $\phi^{(1)} \to 0, \dfrac{\partial \phi^{(1)}}{\partial \eta} \to 0, \dfrac{\partial^2 \phi^{(1)}}{\partial \eta^2} \to 0$ in the unperturbed region $\eta \to \pm\infty$. Hence, we drop the superscript $(1)$ for ease of notation. Thus, the stationary solution to eq. (45) is the standard result,

$$\phi = \phi_m \operatorname{sech}^2\left(\dfrac{\eta}{V}\right),$$

where, the amplitude $\phi_m$ and the width $V$ of the solitary wave are,

$$\phi_m = \dfrac{3U}{A_1},$$

and

$$V = \sqrt{\dfrac{4A_2}{U}}.$$

Fig. 4 illustrates the variation in electron acoustic solitary wave profile within a semiconductor quantum plasma, comparing the cases with and without quantum effects. When quantum effects are present, the soliton profile exhibits a greater amplitude and the width of the soliton is broader, suggesting that the disturbance extends over a larger spatial region. This phenomenon implies that quantum effects contribute to a more extended structure. The Bohm potential introduces a dispersive term proportional to $k^2$ in the quantum hydrodynamic equations. This additional dispersion balances the nonlinearity more effectively, leading to a soliton with a broader width. The inclusion of Fermi pressure, which arises due to electron degeneracy, increases the restoring force within the plasma, which enhances the electrostatic potential, resulting in a larger amplitude soliton.

Fig. 5 illustrates the variation of electron acoustic solitary wave profiles in a semiconductor quantum plasma for different values of the cold-to-hot electron density ratio. As the cold-to-hot electron density ratio increases, the amplitude of the soliton decreases and the soliton becomes broader. The amplitude of a soliton is proportional to the nonlinear coefficient in the KdV equation. As the density of cold electrons increases, the effective charge imbalance between hot and cold electrons is reduced. This reduction weakens the nonlinear steepening effect, which leads to a lower potential peak. The width of the soliton depends on the dispersive coefficient. With an increase in cold electron density, quantum dispersive effects become stronger, allowing the soliton to expand. Since solitons are formed by the balance between nonlinearity and dispersion, a reduction in nonlinearity allows dispersive effects to dominate, resulting in wider solitons.

Fig. 6 illustrates the variation in electron acoustic solitary wave profiles in a semiconductor quantum plasma, comparing scenarios with and without nanoparticle clusters. The solid line, representing the case with nanoparticle clusters, shows a higher peak amplitude than the dashed line, which represents the case without nanoparticle clusters. The increase in amplitude suggests that the presence of nanoparticle clusters enhances the electrostatic potential of the soliton. The soliton in the presence of nanoparticles is slightly broader than the one without nanoparticles. This indicates that the inclusion of nanoparticles increases dispersion effects, leading to a more extended soliton structure.

5. **Summary and Discussion**

In the present paper, the propagation characteristics of electron acoustic waves (EAWs) and solitons in a semiconductor quantum plasma consisting of cold electrons, hot electrons, holes, and ions, with the incorporation of nanoparticle clusters have been studied. The QHD model and KdV approach has been used to establish the dispersion relation and soliton solutions for EAWs, highlighting the impact of nanoparticle doping on the characteristics of electron acoustic modes. Two distinct populations of electrons, low temperature, inertial cold electrons, and high temperature, inertia less hot electrons, along with holes and stationary ions have been considered. The cold electrons provide the inertial contribution to the dynamics, while the hot electrons offer the restoring force. The derived dispersion relation demonstrates how different species contribute to the propagation of EAWs, reflecting the interplay between cold electrons, hot electrons, holes, and nanoparticles. To analyze the nonlinear behaviour of the electron acoustic wave, the perturbation technique has been applied, leading to the KdV equation, which describes the solitary wave solutions.

The analysis show that quantum effects stabilize wave propagation, reducing wave number growth at higher frequencies. The presence of nanoparticle clusters enhances the dispersion properties, leading to significant variations in soliton amplitude and width. The study also demonstrates that, on increasing the cold to hot electron density ratio lower wave number is observed at higher frequencies. The nonlinear analysis through the KdV equation analysis reveals that solitons in quantum plasma exhibit higher amplitudes and broader structures due to quantum diffraction and Fermi pressure. The density ratio of cold to hot electrons affects soliton formation, with an increase in cold electron density leading to lower amplitude and broader solitons due to enhanced quantum dispersion. The inclusion of nanoparticles further enhances the electrostatic potential, leading to an increase in soliton amplitude.

The graphical illustrations further validate the theoretical findings. The dispersion characteristics show distinct differences in the presence and absence of quantum effects, quantum effects stabilize wave propagation, reducing wave number growth at higher frequencies. Similarly, the soliton profiles confirm that the quantum effects and nanoparticle clusters lead to enhanced amplitude and dispersion. These findings have implications for the design and optimization of semiconductor based plasma systems, particularly in high frequency device applications where controlling wave dynamics is crucial.

## Acknowledgement

The authors thank SERB – DST, Govt. of India for financial support under MATRICS scheme (grant no. : MTR/2021/000471).

## References
[1] P.K. Shukla, B. Eliasson, Phys. Usp. 53 (2010) 51;
[2] P.K. Shukla, B. Eliasson, Rev. Mod. Phys. 83 (2011) 885;
[3] S. Ali, W.M. Moslem, P.K. Shukla, R. Schlickeiser, Phys. Plasmas 14 (2007) 082307.
[4] A. Markowich, C. Ringhofer, and C. Schmeiser, Semiconductor Equations (Springer, Vienna, 1990).
[5] G Manfredi, Fields Inst. Commun. 46, 263 (2005)
[6] H. Hang and S. W. Koch, Quantum Theory of the Optial and Electronic Properties of Semiconductors (World Scientific, London, 2004).
[7] W. Li, P. J. Tanner, and T. F. Gallagher, Phys. Rev. Lett. 94, 173001 (2005).
[8] L. K. Ang and P. Zhang, Phys. Rev. Lett. 98, 164802 (2007).
[9] A.P. Misra, Phys. Rev. B 83 (2011) 057401.
[10] W.M. Moslem, I. Zeba, and P.K. Shukla, Appl. Phys. Lett. 101 (2012) 032106.
[11] I. Zeba, M.E. Yahia, P.K. Shukla, and W.M. Moslem, Phys. Lett. A 376 (2012) 2309.
[12] M.E. Yahia, I.M. Azzouz, and W.M. Moslem, Appl. Phys. Lett. 103 (2013) 082105.
[13] Y. Wang and B. Eliasson, Phys. Rev. B 89 (2014) 205316.
[14] Y. Wang and X. Lu, Phys. Plasmas 21 (2014) 022107.
[15] M.R. Amin, Phys. Scr. 90 (2015) 015601.
[16] Moslem WM, Zeba I, Shukla PK. Solitary acoustic pulses in quantum semiconductor plasmas. Appl Phys Lett. 2012;101(1–5):032106.
[17] P.K. Shukla, A. Mamun, B. Eliasson, Geophys. Res. Lett. 31, L07803 (2004)
[18] M. Shalaby, S. El-Labany, R. Sabry, L. El-Sherif, Phys. Plasmas 18, 062305 (2011)
[19] W. El-Taibany, W.M. Moslem, Phys. Plasmas 12, 032307 (2005)

[20] P. Eslami, M. Mottaghizadeh, H.R. Pakzad, Phys. Plasmas 18, 102313 (2011)
[21] N. Dubouloz, R. Pottelette, M. Malingre, G. Holmgren, P.-A. Lindqvist, J. Geophys. Res. 96, 3565 (1991)
[22] N. Dubouloz, R. Treumann, R. Pottelette, M. Malingre, J. Geophys. Res. 98, 17415 (1993)
[23] M Yu and P K Shukla, J. Plasma Phys. 29, 409 (1983)
[24] R L Mace and M A Hellberg, J. Plasma Phys. 43, 239 (1990)
[25] N Dubouloz, R Pottelette, M Malingre and R A Treumann, Geophys. Res. Lett. 18, 155 (1991)
[26] S G Tagare, S V Singh, R V Reddy and G S Lakhina, Nonlinear Process. Geophys. 11, 215 (2004)
[27] E K El-Shewy, Chaos, Solitons and Fractals 31, 1020 (2007)
[28] S A Elwakil, M A Zahran and E K El-Shewy, Phys. Scr. 75, 803 (2007)
[29] I Kourakis and P K Shukla, Phys. Rev. E 69, 036411 (2004)
[30] P K Shukla, L Stenflo and M Hellberg, Phys. Rev. E 66, 027403 (2002)
[31] W. D. Jones, A. Lee, S. M. Gleman, and H. A. Doucet, Phys. Rev. Lett. 35, 1349 (1975).
[32] M. Temerin, K. Cerny, W. Lotko, and F. S. Mozer, Phys. Rev. Lett. 48, 1175 (1982).
[33] Kourakis and P. K. Shukla, Phys. Rev. E 69(3), 036411 (2004).
[34] S. Younsi and M. Tribeche, Astrophys. Space Sci. 330, 295 (2010).
[35] A. Danehkar, N. S. Saini, M. A. Hellberg, and I. Kourakis, Phys. Plasmas 18, 072902 (2011).
[36] M. Shalaby, S. K. El-Labany, R. Sabry, and L. S. El-Sherif, Phys. Plasmas 18, 062305 (2011).
[37] S. Mahmood and W. Masood, Phys. Plasmas 15, 122302 (2008).
[38] O. P. Sah and J. Manta, Phys. Plasmas 16, 032304 (2009).
[39] A. P. Misra, P. K. Shukla, and C. Bhowmik, Phys. Plasmas 14, 082309 2007.
[40] C. Bhowmik, A. P. Misra, and P. K. Shukla, Phys. Plasmas 14, 122107 2007.
[41] W. Masood and A. Mushtaq, Phys. Plasmas 15, 022306 2008.
[42] Mustafa F and Hashim A M 2010 Prog. Electromagn. Res. 104 403
[43] U. Kreibig and M. Vollmer: Optical Properties of Metal Clusters, (Springer, Berlin, 1995).
[44] Kim W, Zide J, Gossard A, Klenov D, Stemmer S, Shakouri A and Majumdar A 2006 Phys. Rev. Lett. 96 045901
[45] Kawasaki J K, Timm R, Delaney K T, Lundgren E, Mikkelsen A and Palmstrom C J 2011 Phys. Rev. Lett. 107 036806
[46] Lung F and Marinescu D C 2011 J. Phys.: Condens. Matter 23 365802
[47] Choudhary K K 2013 J. Nanopart. Res. 15 1362
[48] J. M. Zide, D. O. Klenov, S. Stemmer, A. C. Gossard, G. Zeng, J. E. Bowers, D. Vashaee, A. Shakouri. Appl. Phys. Lett. 87 (2005) 112102.
[49] J. K. Kawasaki, R. Timm, K. T. Delaney, E. Lundgren, A. Mikkelsen, C. J. Palmstrom. Phys. Rev. Lett. 107 (2011) 036806.
[50] W. Kim, J. Zide, A. Gossard, D. Klenov, S. Stemmer, A. Shakouri, A. Majumdar, Phys. Rev. Lett. 96, 045901 (2006).
[51] F. Lung, D.C. Marinescu, J. Phys. Condens. Matter 23, 365802 (2011).
[52] He Z, Zhang Z and Bi S 2020 Mater. Res. Express 7 012004
[53] Wiercigrocha E, Swit P, Kisielewska A, Piwoński I and Malek K 2020 Appl. Surf. Sci. 529 147021

**Figure Captions**

Figure 1. Variation of $\omega/\omega_p$ with $kc/\omega_p$ in semiconductor quantum plasma and in absence of quantum effects $(\hbar = 0)$

Figure 2. Variation of $\omega/\omega_p$ with $kc/\omega_p$ for different value of cold to hot electron density ratio.

Figure 3. Variation of $\omega/\omega_p$ with $kc/\omega_p$ in semiconductor quantum plasma in presence and absence of nanoparticle cluster.

Figure 4. Variation of electron acoustic solitary profiles in semiconductor quantum plasma and in absence of quantum effects $(\hbar=0)$

Figure 5. Variation of electron acoustic solitary profiles for different value of cold to hot electron density ratio.

Figure 6. Variation of electron acoustic solitary profiles in semiconductor quantum plasma in presence and absence of nanoparticle cluster.

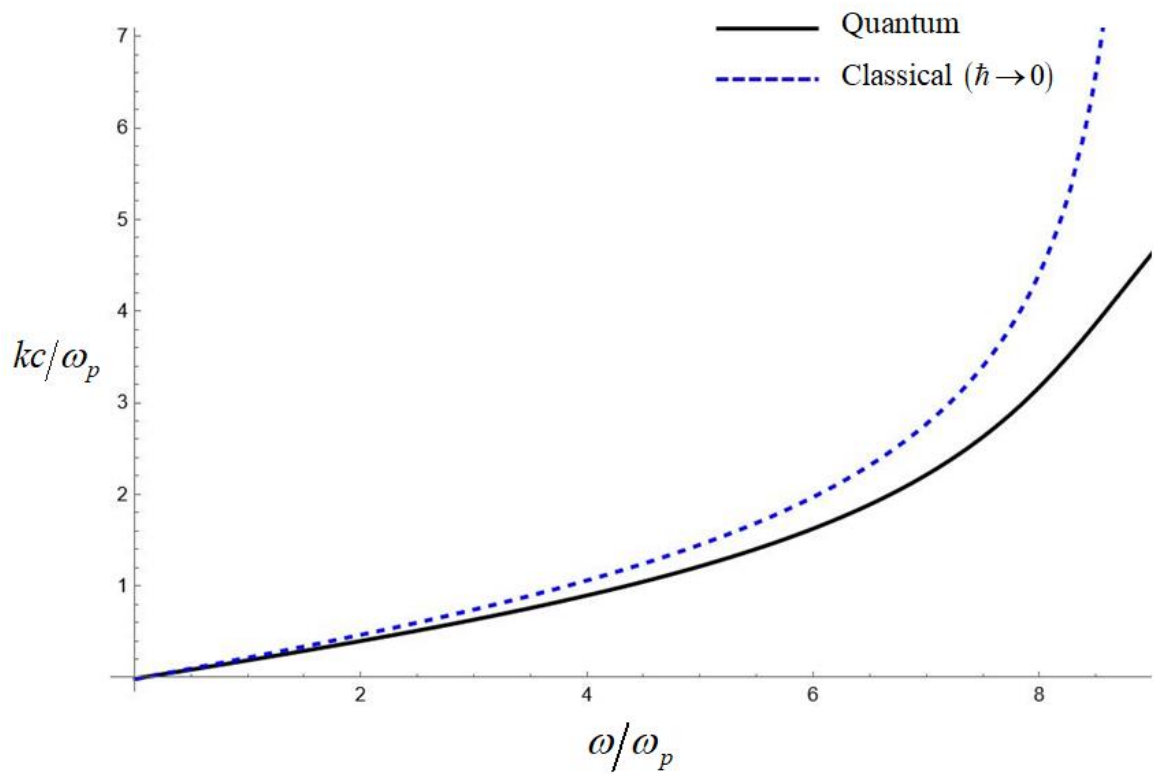

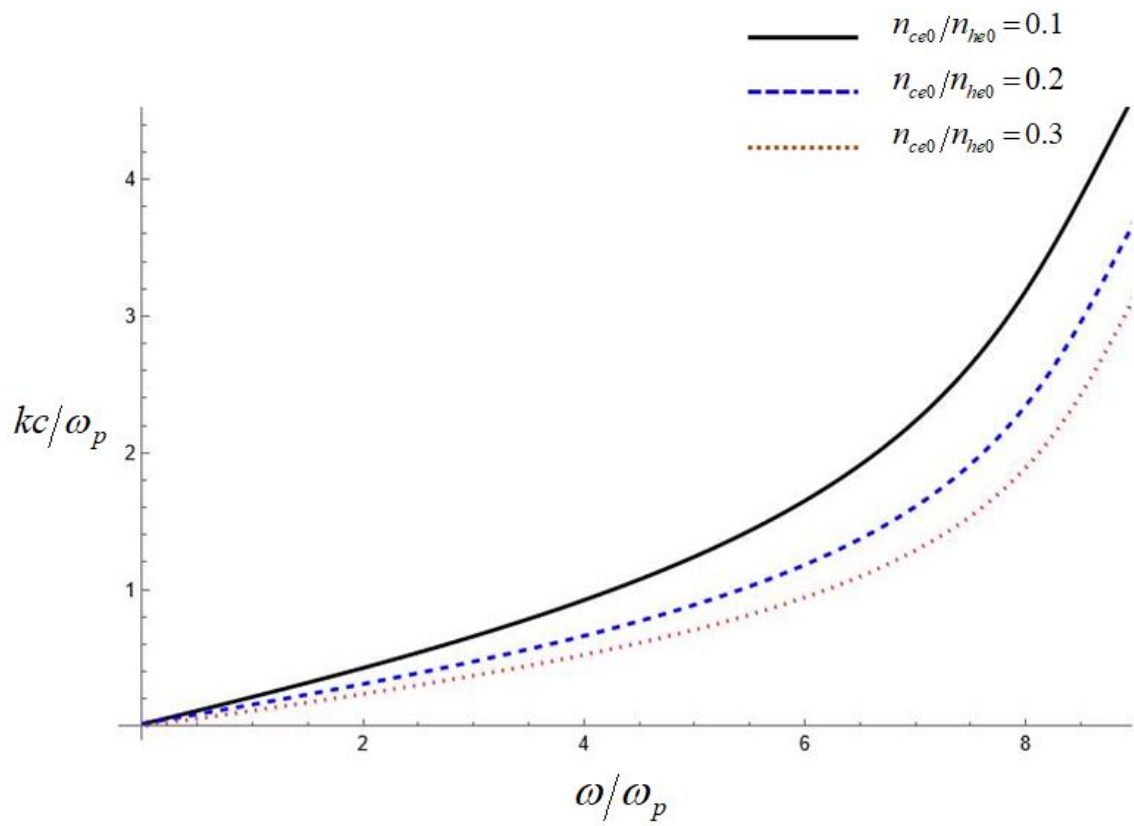

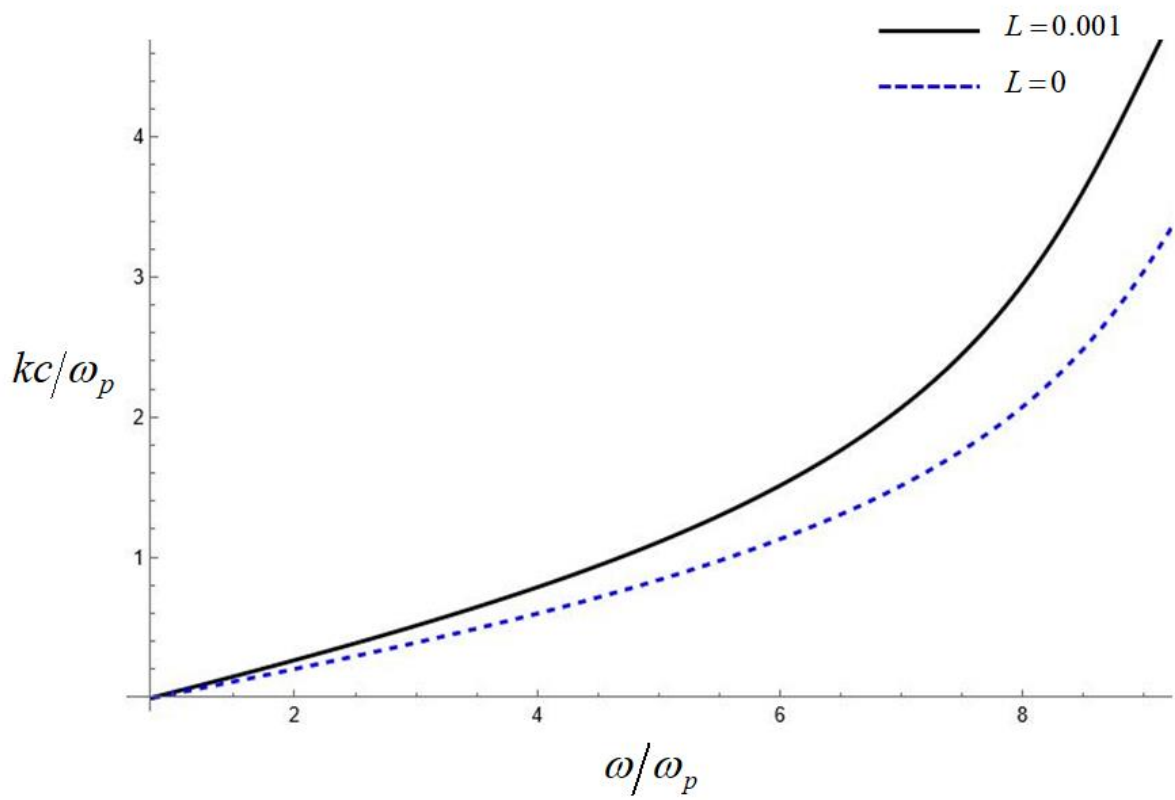

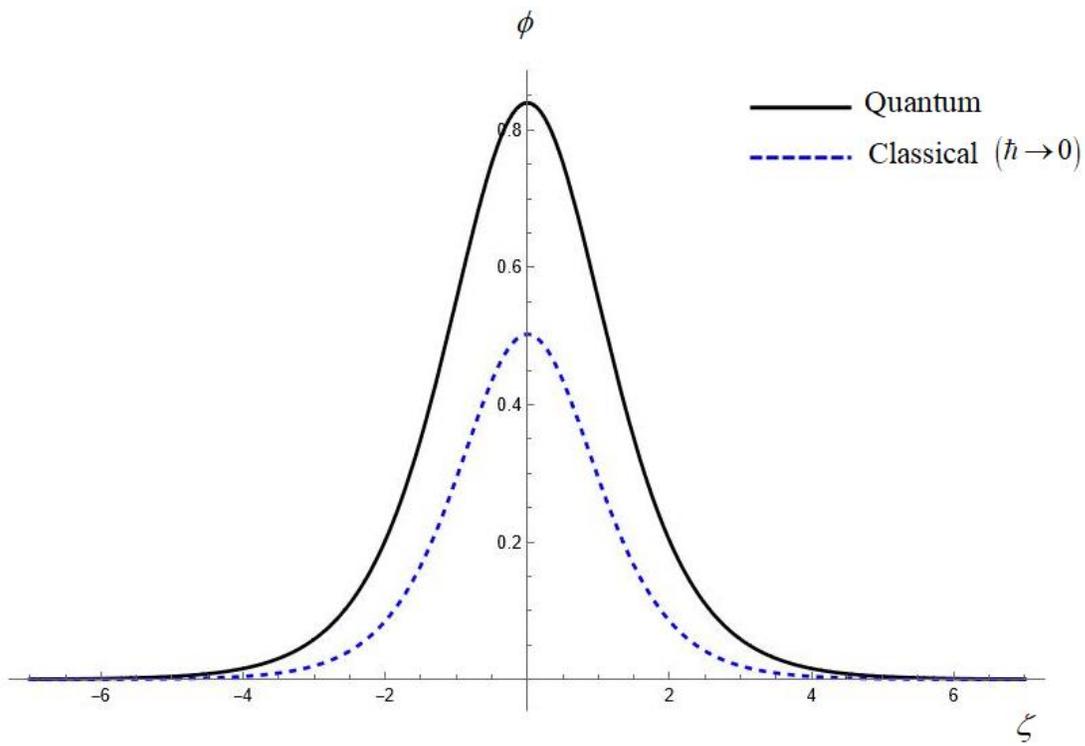

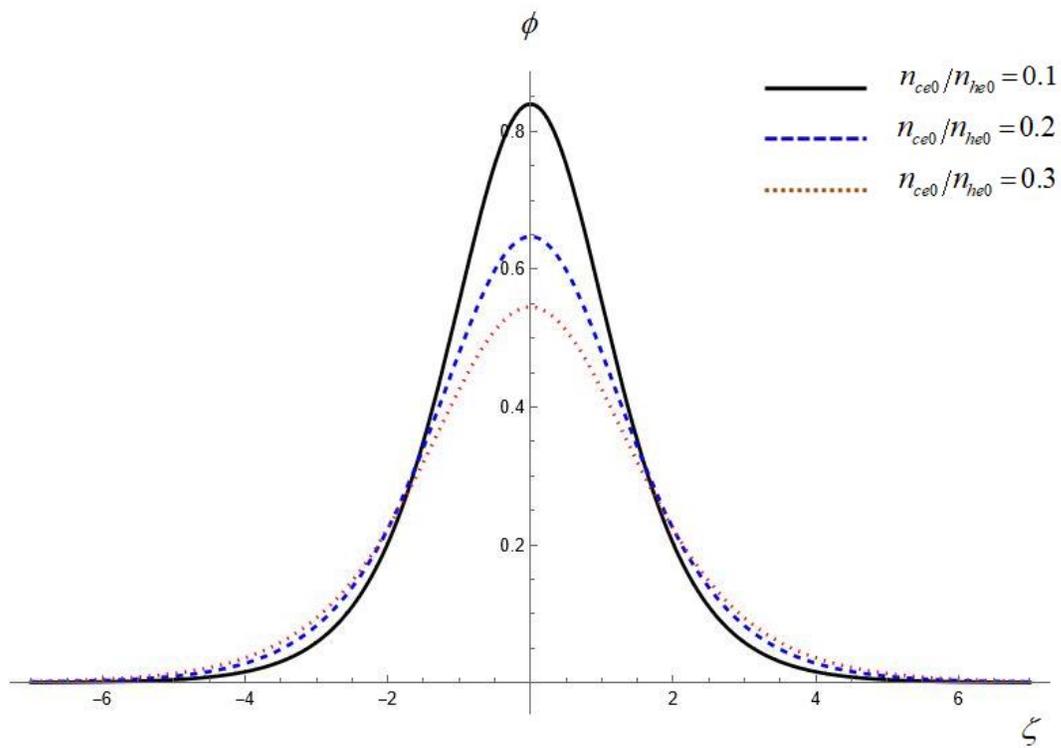

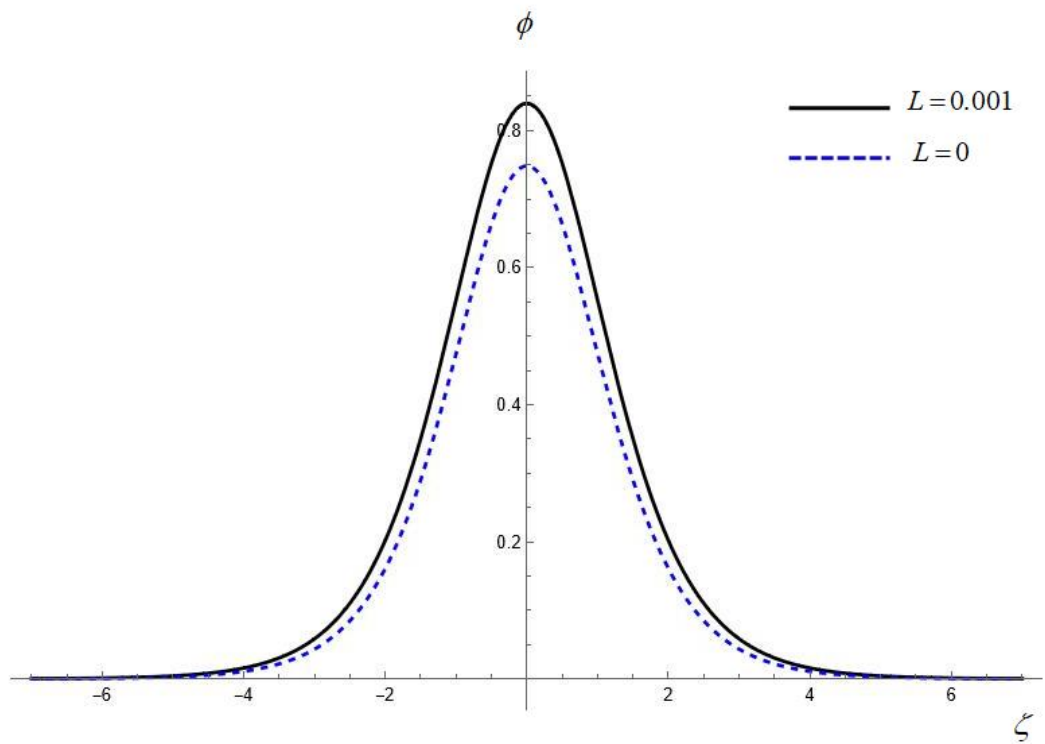